\documentclass[a4paper,12pt]{article}
\usepackage[pctex32]{graphics}

\begin{document}
\topmargin 0pt \oddsidemargin 0mm
\newcommand{\beq}{\begin{equation}}
\newcommand{\eeq}{\end{equation}}
\newcommand{\beqa}{\begin{eqnarray}}
\newcommand{\eeqa}{\end{eqnarray}}
\newcommand{\sr}{\sqrt}
\newcommand{\fr}{\frac}
\newcommand{\mn}{\mu \nu}
\newcommand{\G}{\Gamma}

\begin{titlepage}
\begin{flushright}
INJE-TP-06-02,~hep-th/0603200
\end{flushright}

\vspace{5mm}
\begin{center}
{\Large \bf Phase transitions for the topological  de Sitter
spaces and Schwarzschild-de Sitter black hole} \vspace{12mm}

{\large   Yun Soo Myung \footnote{e-mail
 address: ysmyung@inje.ac.kr}}
 \\
\vspace{10mm} {\em Institute of Mathematical Science and School of
Computer Aided Science \\ Inje University, Gimhae 621-749, Korea}
\end{center}
\vspace{5mm} \centerline{{\bf{Abstract}}}
 \vspace{5mm}
We study whether  the Hawking-Page phase  transition may occur in
 topological de Sitter
 spaces (TdS) and Schwarzschild-de Sitter black hole (SdS).
 We show that at the critical temperature $T=T_1$, TdS with  hyperbolic cosmological horizon can make
 the
 Hawking-Page transition from the zero mass de Sitter space  to TdS.
It is also shown that there is no Hawking-Page  transition for TdS
with Ricci-flat and spherical horizons, when the zero mass  de
Sitter space is taken as the thermal background. Also we find that
the SdS undergoes a different  phase transition at $T=0$ which the
Nariai black hole is formed. Finally we connect our results to the
dS/CFT correspondence.
\end{titlepage}
\newpage
\renewcommand{\thefootnote}{\arabic{footnote}}
\setcounter{footnote}{0} \setcounter{page}{2}

\section{Introduction}
A number of authors have  shown that for a large class of black
holes, the Bekenstein-Hawking entropy receives logarithmic
corrections due to thermodynamic fluctuations~\cite{KM1}. A
corrected formula takes the form \beq \label{CEN} S_c=S-\fr{1}{2}
\ln C + \cdots, \eeq where $C$ is the specific heat of the given
system,  and $S$ denotes the uncorrected Bekenstein-Hawking
entropy. Here  $C$ should be positive for  Eq.(\ref{CEN}) to be
well-defined. We note that for $C>0~(C<0)$, the system is
thermodynamically stable (unstable).  A black hole with negative
specific heat is  in an unstable equilibrium with the  heat
reservoir of the temperature $T$ \cite{GPY}. Its fate under small
fluctuations will be either to decay to  hot flat space or to grow
without limit by absorbing thermal radiation in the  heat
reservoir~\cite{York}. There exists a way to achieve a stable
black hole in an equilibrium with the heat reservoir. A black hole
could be rendered thermodynamically stable by placing it in AdS
space. An important point is to understand how a black hole with
positive specific heat could emerge from thermal radiation through
a phase transition. To this end, one introduces the Hawking-Page
phase transition between thermal AdS space and Schwarzschild-AdS
black hole~\cite{HP,BCM,Witt}.

Further, there is  a close similarity between the event horizon of
a black hole  and the cosmological horizon of de Sitter
space~\cite{GH}. Hence, it is interesting to study the thermal
properties of various de Sitter spaces.
 In this work, we check whether a  Hawking-Page phase transition occurs in topological
de Sitter spaces. Also we study the phase transition of the
Schwarzschild-de Sitter black hole, where a black hole is inside
the cosmological horizon. Moreover, we investigate the
implications of Hawking-Page transition on dS/CFT correspondence.

Our study is based on the observations of heat capacity, free
energy, and generalized (off-shell) free energy near the phase
transition.

The organization of this work is as follows. Section 2 is devoted
to reviewing topological AdS black holes. We discuss the
thermodynamic properties of  topological de Sitter spaces which
are similar to those of  topological AdS black holes in section 3.
The thermodynamic  properties of Schwarzschild-de Sitter black
hole are investigated  in section 4. In section 5 we study whether
or not the Hawking-Page phase transition occurs in TdS and SdS. We
connect our results to the dS/CFT correspondence in section 6 by
introducing boundary CFT and Cardy-Verlinde formula. Finally, we
discuss our results in section 7.

\section{Topological AdS black holes}
\begin{figure}[t!]
 \centering
   \includegraphics{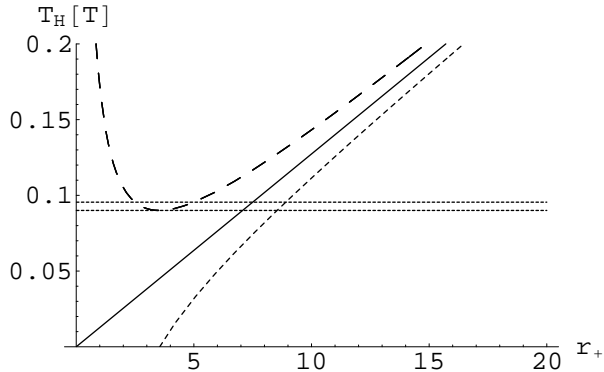}
    \caption{Temperature for TAdS with $r_+=r_E$.
 The long-dashed curve denotes  $T_H^{SAdS}(\ell=5,r_+)$, showing a
minimum temperature $T=T_0$ at $r_+=r_0=3.5$ with units $G_5=1$
and $\ell=5$. The solid line is a  monotonically increasing
function $T_H^{FAdS}(\ell=5,r_+)$. The short-dashed curve
represents $T_H^{HAdS}(\ell=5,r_+\ge r_0)$, indicating a forbidden
region of $0\le r_+ <r_0$. Two horizontal lines denote the
temperatures $T=T_0$ and $T_1$ for the heat reservoir. }
\label{fig1}
\end{figure}

We begin with a review of topological black holes in AdS space
\cite{BIR}, as these are thermodynamically related to TdS
\cite{Myu}. A black hole in asymptotically flat spacetime has  a
spherical horizon. If one introduces a negative cosmological
constant, a black hole  can have a non-spherical horizon. This is
called  the topological AdS black holes (TAdS) whose  metric  in
five dimensions is given by
 \beq ds^{2}_{TAdS}=
 -h(r)dt^2 +\fr{1}{h(r)}dr^2 +r^2d\Sigma^2_k,
\label{BMT} \eeq where $d\Sigma^2_k= d\chi^2
+f_{k}(\chi)^2(d\theta^2+ \sin^2 \theta d\phi^2)$ describes the
horizon geometry with a constant curvature. Further $h(r)$ and
$f_k(\chi)$ are given by \beq h(r)=k-\fr{m}{r^2}+ \fr{
r^2}{\ell^2},~~~ f_{0}(\chi) =\chi, ~f_{1}(\chi) =\sin \chi,
~f_{-1}(\chi) =\sinh \chi. \eeq Here we define $k$=1,~0, and $-1$
cases as the Schwarzschild-AdS black hole (SAdS)~\cite{MP,CMu},
flat-AdS  black hole (FAdS), and hyperbolic-AdS  black hole
(HAdS)~\cite{CAI1}, respectively. In the case of $k=1$ and $m=0$,
we have a thermal AdS space with curvature radius $\ell$.
\begin{figure}[t!]
 \centering
   \includegraphics{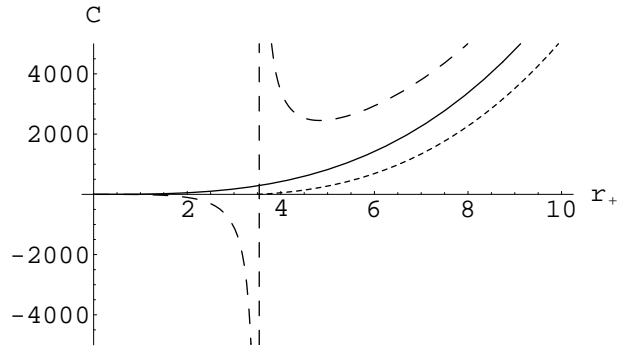}
    \caption{Specific heats for TAdS.
     Here $r_+=r_E$ plays the role of effective
    temperature.
 The
long-dashed curve denotes  $C^{SAdS}(\ell=5,r_+)$, and diverges at
$r_+=r_0$. The solid curve represents a monotonically increasing
specific heat $C^{FAdS}(\ell=5,r_+)$. The short-dashed curve
indicates $C^{HAdS}(\ell=5,r_+\ge r_0)$, which shows  a forbidden
region of $0\le r_+ <r_0$. } \label{fig2}
\end{figure}
\begin{figure}[t!]
 \centering
   \includegraphics{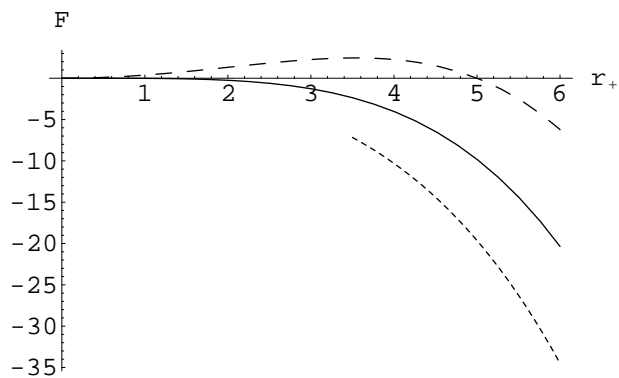}
    \caption{Three graphs of free energy for TAdS.
The long-dashed curve represents $F^{SAdS}(\ell=5,r_+)$, which
shows a continuous transition from positive value to negative one
at $r_+=r_1$.
   The solid curve represents   $F^{FAdS}(\ell=5,r_+)$,  while the
short-dashed  curve denotes  $F^{HAdS}(\ell=5,r_+\ge r_0)$ with a
forbidden region. } \label{fig3}
\end{figure}

The location of the  event horizon is given by
 \beq \label{EH} r_{E}^2=
\fr{\ell^2}{2}\Big(-k+\sqrt{k^2 +4 m/\ell^2}\Big). \eeq The
relevant thermodynamic quantities: reduced mass ($m$), free energy
($F$), Bekenstein-Hawking entropy ($S$), Hawking temperature
($T_H$),  energy (ADM mass: $E=F+T_H S=M$), and specific heat
($C$) are given by\footnote{In defining the energy $E$ and free
energy $F$ for the HAdS of $k=-1$, we do not include $M_{\rm
crit}\equiv M|_{r_+=r_0}=-3V_3\ell^2/64 \pi G_5$ of
Ref.\cite{BIR}. Even if we include this term, the phase transition
is unaffected. This is because it is a constant~\cite{myung3}.}
\beqa \label{TQ} &&m=r_{E}^2 \Big(\fr{r_{E}^2}{\ell^2}+ k
\Big),~~F=-\fr{V_3 r_{E}^2}{16 \pi G_5}\Big(\fr{r_{E}^2}{\ell^2}-
k \Big),~~ S=\fr{V_3r_{E}^3}{4G_5},~~\\ \nonumber &&T_H=\fr{k}{2
\pi r_{E}} +\fr{r_{E}}{\pi \ell^2},~~E=\fr{3V_3m}{16 \pi G_5},~~C=
3 \fr{2r_{E}^2+k\ell^2}{2r_{E}^2-k\ell^2}S, \eeqa where $V_3$ is
the volume of a unit three-dimensional hypersurface $\Sigma_k$ and
$G_5$ is the five-dimensional Newton constant. The different
behaviors of temperature are shown in Fig. 1~\cite{CTZ}. From the
equilibrium condition ($T=T_H$) of a black hole with the  heat
reservoir, we find the two solutions: a small, unstable black hole
of  size \beq \label{UBH} r_u =\frac{\pi
\ell^2T}{2}\Bigg[1-\sqrt{1-\frac{8k}{(2\pi \ell T)^2}}\Bigg] \eeq
and a large, stable black hole of  size \beq \label{SBH}
r_s=\frac{\pi \ell^2T}{2}\Bigg[1+\sqrt{1-\frac{8k}{(2\pi \ell
T)^2}}\Bigg].\eeq Here  we find that $r_u=0,~r_s=\pi \ell^2T$ for
$k=0$; $r_u \simeq 1/2\pi T,~r_s \simeq \pi \ell^2T$ for $k=1$ and
$T \gg 1/\ell$; $r_s \simeq \pi \ell^2T$ for $k=-1$ and $T \gg
1/\ell$. For $k=1$,  an extremum  is at
$r_u=r_s=r_0=\ell/\sqrt{2}$ and the corresponding temperature is
given by  $T_0=\sqrt{2}/\pi \ell$. Further,  we obtain a critical
point $r_1=\ell$  from the condition of  $F=0$. The corresponding
temperature is given by $T_1=3/2\pi \ell$. At this temperature, we
have $r_u=\ell/2$ and $r_s=\ell$. As is shown in Fig. 2, there are
three different specific heats. $C^{SAdS}$ has a simple pole at
$r_+=r_0$, while $C^{FAdS}$ is a continuous increasing function.
In contrast, $C^{HAdS}$ has a forbidden region which leads to the
interruption of thermodynamic analysis. Finally, we observe  from
Fig. 3 that $F^{FAdS}$ and $F^{HAdS}$ are monotonically decreasing
functions, but  $F^{HAdS}$ has a forbidden region. $F^{SAdS}$ has
a maximum value at $r_+=r_0$ and is zero at $r_+=r_1$, which shows
a feature of the Hawking-Page transition.

\section{Topological de Sitter spaces}

The topological de Sitter space (TdS) solution was originally
introduced to check the mass bound conjecture in de Sitter space:
any asymptotically de Sitter space with the mass greater than an
exact de Sitter space has a cosmological singularity~\cite{TDS}.
It is very interesting to study the thermodynamic properties of
this kind of cosmological horizon.  We consider the topological de
Sitter solution in five-dimensional spacetime \beq ds^{2}_{TdS}=
-h(r)dt^2 +\fr{1}{h(r)}dr^2 +r^2 d\Sigma^2_k, \label{Tds} \eeq
where $k=0,~\pm1$ and  $h(r)$ is  given by \beq
h(r)=k+\fr{m}{r^2}- \fr{ r^2}{\ell^2}. \eeq Requiring $m>0$ leads
to the fact that  the event horizon of  black hole disappears and
instead, a naked singularity appears at $r=0$ inside the
cosmological horizon. Here we define $k=1,~0,$ and $-1$ cases as
the Schwarzschild-topological de Sitter
 space (STdS), flat-topological de Sitter  space (FTdS), and
hyperbolic--topological de Sitter  space (HTdS), respectively. In
the case of $k=1$ and $m=0$, we have an exact de Sitter space with
 curvature radius $\ell$.
The  cosmological horizon is at  \beq \label{CH} r_{C}^2=
\fr{\ell^2}{2}\Big(k+\sqrt{k^2 +4 m/\ell^2} \Big). \eeq  We note
that there is no restriction on $r_C$, in contrast with the
cosmological horizon of SdS. The thermodynamic quantities for the
cosmological horizon are ~\cite{CAI2}\beqa \label{3TQ} &&m=r_{C}^2
\Big(\fr{r_{C}^2}{\ell^2}- k \Big),~~F=-\fr{V_3 r_{C}^2}{16 \pi
G_5}\Big(\fr{r_{C}^2}{\ell^2}+ k \Big),~~
S=\fr{V_3r_{C}^3}{4G_5},~~\\ \nonumber &&T_H=-\fr{k}{2 \pi r_{C}}
+\fr{r_{C}}{\pi \ell^2},~~E=\fr{3V_3m}{16 \pi G_5}=M,~~C= 3
\fr{2r_{C}^2-k\ell^2}{2r_{C}^2+k\ell^2}S, \eeqa where $V_3$ is the
volume of a unit three-dimensional hypersurface $\Sigma_k$.
Considering the
 equilibrium condition  $T=T_H$, we find  that two
solutions: a small, unstable cosmological horizon of  size \beq
\label{UCH} r_u =\frac{\pi
\ell^2T}{2}\Bigg[1-\sqrt{1+\frac{8k}{(2\pi \ell T)^2}}\Bigg]\eeq
and a large, stable cosmological horizon of  size \beq \label{SCH}
r_s=\frac{\pi \ell^2T}{2}\Bigg[1+\sqrt{1+\frac{8k}{(2\pi \ell
T)^2}}\Bigg].\eeq From the above, we find $r_u=0,~r_s=\pi \ell^2T$
for $k=0$; $r_u \simeq 1/2\pi T,~r_s \simeq \pi \ell^2T$ for
$k=-1$ and $T \gg 1/\ell$; $r_s \simeq \pi \ell^2T$ for $k=1$ and
$T \gg 1/\ell$. For $k=-1$, there is  an extremum at
$r_u=r_s=r_0=\ell/\sqrt{2}$ and $T_0=\sqrt{2}/\pi \ell$. Further,
 we obtain a critical point $r_1=\ell$ from the condition of $F=0$. The
corresponding temperature is also given by $T_1=3/2\pi \ell$.

All results of TdS solution may be recovered from TAdS  by
substituting  $k$ and $r_E$ into $-k$ and $r_C$: SAdS $\to $ HTdS,
HAdS $\to$ STdS, and  FAdS $\to$ FTdS. Then we may obtain all
thermodynamic behaviors of TdS from Figs. 1, 2 and 3.

\begin{figure}[t!]
 \centering
   \includegraphics{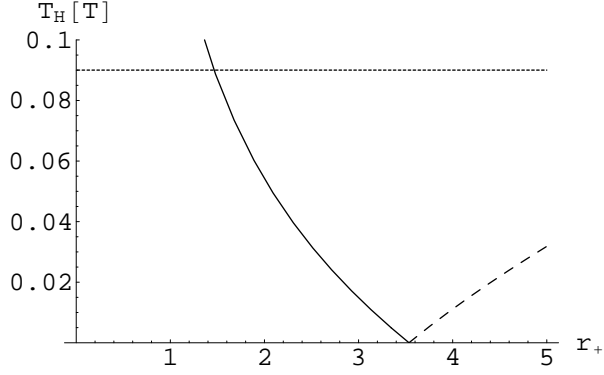}
    \caption{Temperature for SdS.
    Here $r_+$ represents $r_E$ for the event horizon and $r_C$
    for the cosmological horizon. Here $r_E~(r_C)$ is confined to $0<r_E \le r_0~(r_0 \le
    r_C <\ell)$.
     The solid curve represents the temperature of the event horizon $T^{ESdS}_H$,
     while the dashed curve denotes  the temperature of the cosmological
     horizon $T^{CSdS}_H$. The horizontal line represents the temperature $T=T_0$ of the
    heat bath.} \label{fig4}
\end{figure}

\begin{figure}[t!]
 \centering
   \includegraphics{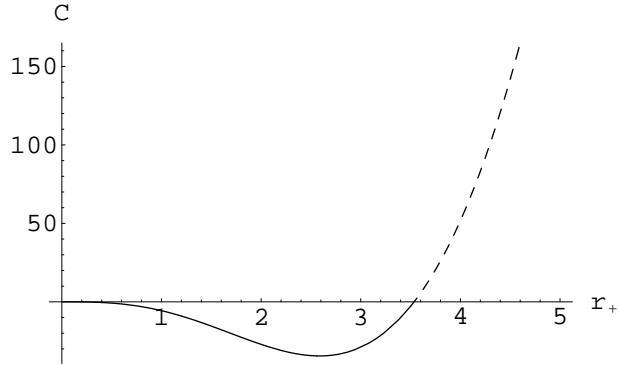}
    \caption{The graph of specific heat $C^{ESdS}$ and $C^{CSdS}$ for  SdS.
    Here $r_+$ represents $r_E$ for the event horizon and $r_C$
    for the cosmological horizon.  } \label{fig5}
\end{figure}

\section{ Schwarzschild-de Sitter black hole}

 \begin{figure}[t!]
 \centering
   \includegraphics{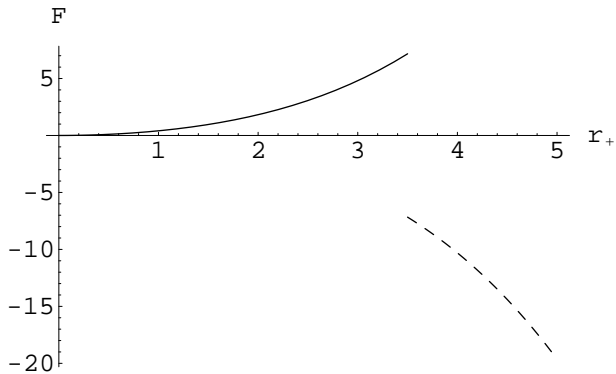}
    \caption{The plot for  free energy for  SdS.
    The solid curve represents the  free energy $F^{ESdS}$ for the event horizon,
    while the dashed curve denotes the free energy $F^{CSdS}$ for the  cosmological horizon.
    There exists a discontinuity at $r_+=r_0$ due to the energy gab.
    At this point the Nariai black hole is formed.  } \label{fig6}
\end{figure}
It seems that there is no essential difference between TAdS and
TdS thermodynamically. Therefore, we are curious to study  the
thermal property of a black hole in de Sitter space. For this
purpose,  we introduce the Schwarzschild-de Sitter  black hole in
five-dimensional spacetime~\cite{CAI2} \beq ds^{2}_{SdS}=
-h(r)dt^2 +\fr{1}{h(r)}dr^2 +r^2 d\Omega_3^2, \label{SDS} \eeq
where $h(r)$ is given by \beq h(r)=1-\fr{m}{r^2}- \fr{
r^2}{\ell^2}.\eeq
 In
the case of $m=0$, we have an exact de Sitter space with the
largest cosmological horizon ($r_C=\ell$). However, $m \not=0$
generates the SdS black hole.
 The cosmological and event horizons
are located at  \beq \label{3EH} r_{C/E}^2=
\fr{\ell^2}{2}\Big(1\pm \sqrt{1 -4 m/\ell^2}\Big). \eeq We
classify three cases with $r_0=\ell/\sqrt{2}$:  $\sqrt{2m}=r_0$,
$\sqrt{2m}>r_0$, and  $\sqrt{2m}<r_0$. The case of $\sqrt{2m}=r_0$
corresponds to the maximum black hole and the minimum cosmological
horizon (that is, Nariai black hole). Here we have
$r_{E}=r_{C}=r_0$. A large black hole  of $\sqrt{2m}>r_0$ is not
allowed  in de Sitter space. The case of $\sqrt{2m}<r_0$
corresponds to a small black hole inside the cosmological horizon.
In this case a cosmological horizon is located  at $r=r_{C}$ with
$r_{C} \simeq \ell\sqrt{1-m/\ell^2}$, while  an  event horizon is
at $r=r_{E}$ with $r_{E} \simeq \sqrt{m}$. Hence, restrictions on
$r_E$ and $r_C$  are given by  \beq \label{2Ineq} 0< r_{E} \le
r_0,~~ r_0 \le r_{C} <\ell. \eeq As the reduced mass $m$
approaches  the maximum value of $m=\ell^2/4$, the small black
hole increases to the Nariai black hole at $r_E=r_0$, whereas the
cosmological horizon decreases  to the minimum value of $r_C=r_0$.
This property  contrasts to the TdS solution: there is no upper
limit on the size $r_C$ of cosmological horizon in topological de
Sitter spaces.

 The thermodynamic quantities for two horizons   are given
by~\cite{CM,NOO} \beqa \label{2TQ} &&m_{E/C}=r_{E/C}^2
\Big(-\fr{r_{E/C}^2}{\ell^2}+ 1 \Big),~~F_{E/C}=\pm\fr{V_3
r_{E/C}^2}{16 \pi G_5}\Big(\fr{r_{E/C}^2}{\ell^2}+1 \Big),~~
S_{E/C}=\fr{V_3r_{E/C}^3}{4G_5},~~\\ \nonumber
&&T_H^{E/C}=\pm\fr{1}{2 \pi r_{E/C}} \mp \fr{r_{E/C}}{\pi
\ell^2},~~E_{E/C} =\pm \fr{3V_3m}{16 \pi G_5},~~C_{E/C}= 3
\fr{2r_{E/C}^2-\ell^2}{2r_{E/C}^2+\ell^2}S_{E/C}, \eeqa where
again $V_3$ is the volume of a unit three-dimensional sphere
$\Omega_3$. We note here that there exists an energy gab $\Delta
E=E_E-E_C$ at $r_+=r_0$ between two horizons. This gives rise to
the discontinuity in free energy.

Using the equilibrium condition $T=T_H$, we obtain a small,
unstable black hole of  size \beq \label{ubhs} r_u =\frac{\pi
\ell^2T}{2}\Bigg[-1+\sqrt{1+\frac{8}{(2\pi \ell T)^2}}\Bigg]\eeq
and a large, stable cosmological horizon of  size \beq
\label{sbhs} r_s=\frac{\pi
\ell^2T}{2}\Bigg[1+\sqrt{1+\frac{8}{(2\pi \ell T)^2}}\Bigg].\eeq
However, considering the restrictions in Eq.(\ref{2Ineq}) together
with $T\gg 1/\ell$, $r_s>\ell$ is not allowed in the SdS. Hence
the two temperatures of $T_0$ and $T_1$  appeared in TAdS and TdS
may  not be  relevant to the SdS phase transition.

As is shown in Fig. 4, the temperature $T^{E/C}_H$ behaves
differently from   the TAdS and TdS case. At $T=T_0$, we read off
an unstable black hole of  size $r_u=1.46$ from a junction of line
and curve but there is no stable cosmological horizon. Making use
of Eqs. (\ref{2Ineq}) and (\ref{2TQ}), one always finds a negative
specific heat for the event horizon (ESdS) and a positive specific
heat for the cosmological horizon (CSdS). See Fig. 5. The solid
curve represents a negative specific heat $C^{ESdS}(\ell=5,0< r_+
\le r_0)$
   for the event horizon,
    which shows the thermal
   instability clearly.
   At $r_+=r_0$, we have the Nariai black hole with $T_H=0$. The
 dashed curve denotes the positive specific heat $C^{CSdS}(\ell=5,r_0 \le r_+ < \ell)$
 for the cosmological
  horizon, indicating the thermal stability.

Further, it is observed  from Fig. 6 that one cannot obtain any
critical point of  $r_+=r_1$ from the condition of  $F_{E/C}=0$.
Instead we find a discontinuity $\Delta F=3(5/4)^2 \pi$ at
$r_+=r_0$, which shows that a phase transition may occur between
the  event and cosmological horizon. This arises because  the
energy gab exists between event and cosmological horizons. As is
shown in Fig. 4, we note that $T_H \to 0$, as $r_+$ approaches
$r_0$. This temperature may be a candidate for the critical
temperature at the ESdS-CSdS phase transition.

\section{Hawking-Page phase transition}

We start with  the known case of SAdS.  In order to understand the
Hawking-Page phase transition, we introduce the generalized free
energy~\cite{CMu} \beq \label{8eq4} F^{off}(\ell,r_+,T)=
E(\ell,r_+)- T\cdot S(r_+) \eeq which applies to any value of
$r_+$ with a fixed temperature $T$. As is shown in Fig. 7, for
$T=T_0$, an extremum appears at $r_+=r_0(=r_u=r_s)$. We confirm
that for $T>T_0$, there are two saddle points: a small, unstable
black hole of  radius $r_u$ and a large, stable black hole of
radius $r_s$. $F$ is composed of  a set of two saddle points for
$F^{off}$. That is, $F$ can be obtained from $F^{off}$ through
operation of finding saddle points: $\partial F^{off}/\partial
r_+=0 \to T=T_H \to F=F^{off}|_{T=T_H}$. Hence, the free energy
$F$ is called  the on-shell (equilibrium) free energy, whereas the
generalized free energy $F^{off}$ corresponds to the off-shell
(non-equilibrium) free energy.
\begin{figure}[t!]
 \centering
   \includegraphics{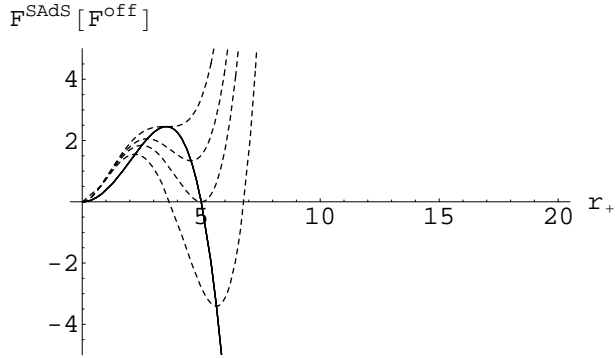}
    \caption{The process of a black hole nucleation known as  the Hawking-Page phase transition.
   The solid curve represents the free
energy $F^{SAdS}(\ell=5,r_+)$,  while the dashed curves denote the
generalized free energy $F^{off}(\ell=5,r_+,T)$ for four different
temperatures: from  top to bottom, $T=T_0(=0.09),~ 0.093,~
T_1(=0.095), $ and  0.1.} \label{fig7}
\end{figure}

At this stage, we briefly review the Hawking-Page phase transition
\cite{HP}. For $T_0<T<T_1$, we find  the  inequality
 by noting  the second dashed graph in Fig. 7 \beq
\label{9eq4} F^{off}(r_+=0)<F^{off}(r_+=r_s)
<F^{off}(r_+=r_u),\eeq
 which means that the saddle point at $r_+=0$
(thermal AdS) dominates. For $T>T_1$, the large, stable black hole
dominates because $F^{off}(r_+=r_s)<0$. Actually, there is a
change of dominance at the critical temperature $T=T_1$. In the
case of $T<T_1$, the system is described by a thermal gas, whereas
for $T>T_1$, it is described by a large, stable black hole. In
this case, the small, unstable  black hole plays a crucial role as
the mediator for the transition between thermal AdS and a large,
stable black hole.  This is the Hawking-Page transition for a
black hole nucleation in AdS space~\cite{BCM}. The same phase
transition may occur between the zero mass de Sitter space and
HTdS.

 For the FAdS (FTdS) case, the small, unstable black hole is not
present but the stable black hole is present for any $T>0$. Hence
the transition from $M=0~(r_+=0)$ to $M\not=0~(r_+\not=0)$ is
possible by an off-shell processes. This process is described by
the off-shell free energy $F^{off}$. As is shown Fig. 8, at $T=0$,
the thermal AdS ($r_+=0$) is more favorable than $r_+ \not=0$. The
two saddle points correspond to the endpoints of the transition
for $T=T_0$ and $T_1$. However, this does not belong to the
 Hawking-Page transition because there is no
mediator~\cite{myung3}. In this case, $r_u=0$ plays the role of a
saddle point for the thermal AdS space. The same off-shell
transition may occur  between the zero mass de Sitter space and
FTdS.

Concerning  the HAdS (FTdS) case, any  phase transition  is
unlikely to occur.
 This is
because of the forbidden region $0\le r_+ <r_0$, which makes an
difficulty to define the temperature $T_H$  and the radius of
unstable black hole $r_u$. Therefore, we could not carry out a
complete analysis for the HADS and STdS.
\begin{figure}[t!]
 \centering
   \includegraphics{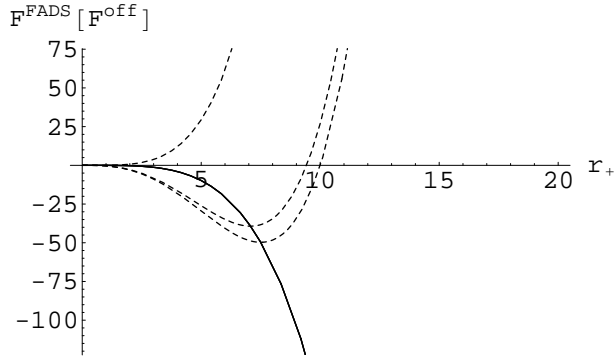}
    \caption{The off-shell transition  for FAdS without a small, unstable black hole.
   The solid curve  represents the free
energy $F^{FAdS}(\ell=5,r_+)$,  while the dashed curves denote the
generalized free energy $F^{off}(\ell=5,r_+,T)$ for three
different temperatures: from top to bottom $T=0,~ T_0(=0.09),$ and
$T_1(=0.095)$.  The stable black holes exist as saddle points at
$r_s=7.07~ (T=T_0)$ and $r_s=7.5 ~(T=T_1)$. } \label{fig8}
\end{figure}

\begin{figure}[t!]
 \centering
   \includegraphics{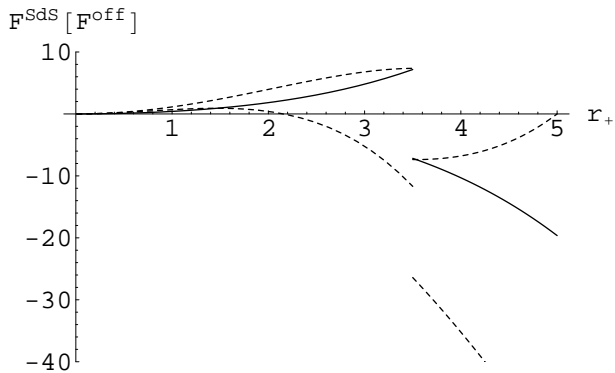}
    \caption{The assumed transition process for SdS.
   The solid lines represent the free
energies $F^{ESdS}(\ell=5,0 < r_+ \le r_0)$ and $F^{CSdS}(\ell=5,
r_0 \le r_+ <\ell)$. The dashed lines at top denote the
generalized free energies $F^{off}(\ell=5,0<r_+ \le r_0,T=0)$ and
$F^{off}(\ell=5,r_0\le r_+ \le \ell,T=0)$. The dashed lines at
bottom show the generalized free energies at $T=T_0$. There exist
discontinuities  at $r_+=r_0$ due to the energy gab between two
horizons. } \label{fig9}
\end{figure}

Now we are in a position to  discuss the SdS transition.  First,
we assume that there is a phase transition between the ESdS and
CSdS at $T=T_0$, as is suggested by the SAdS (HTdS). As is shown
in Fig. 9, at $T=T_0$, we obtain the location $r_u=1.46$ of the
unstable black hole from the condition of  $F^{ESdS}=F^{off}$.
Also we confirm this location from Fig. 4.  However, the stable
black hole is not present here. We note that the allowed region is
confined to $0\le r_+ \le \ell$. In this case, the stable black
hole is located at $r_s=8.5$ which is beyond $r_+=\ell$. Also
there exist a discontinuity at $r_+=r_0$, which may be an obstacle
to interpreting this process as a phase transition.  As a result,
this transition  occur unlikely at $T=T_0$. Similarly,  choosing
$T=T_1$ leads to the similar result.

Hence, we would like to introduce  the Nariai phase  transition at
$T=0$. As was emphasized previously, the Nariai black hole has
zero temperature. This is a case of the maximum black hole and the
minimum cosmological horizon. In Fig. 9, we have
$F^{ESdS}=F^{off}$ at $r_u=r_0$ and $F^{CSdS}=F^{off}$ at
$r_s=r_0$. This means that the location $r_+=r_0$ is not only the
critical point of phase transition  but also the position of the
stable cosmological horizon. This arises because of a peculiar
property of the reduced mass $m$ in de Sitter space. As $m$
increases to its  maximum value ($m=\ell^2/4$), $r_E$ increases to
the radius of the Nariai black hole, whereas the cosmological
horizon decreases to the minimum ($r_C=r_0$). We observe  that at
$T=0$, $r_C=r_0$ is more favorable than $r_C \not= r_0$.

Finally, we comment on the SAdS transition that the free energies
are the same at $r_+=r_0(=r_u=r_s)$:
$F^{SAdS}(\ell=5,r_u)=F^{SAdS}(\ell=5,r_s)$ from Fig. 7, but the
specific heat $C$ changes from $-\infty$ to $\infty$ at $r_+=r_0$
from Fig. 2. On the other hand,  we observe for the Nariai
transition that $F^{ESdS}(\ell=5,r_u)\not=F^{CSdS}(\ell=5,r_s)$,
but $C^{ESdS}(\ell=5,r_u)=C^{CSdS}(\ell=5,r_s)$. Hence, the Nariai
transition is not the Hawking-Page transition. However, the
interpretation of Nariai transition may be incorrect because we
have two horizons of $r_E$ and $r_C$. The Killing vector
$K=\partial_t$ is timelike in region $r_E<r<r_C$, while it is
spacelike in the others of $0<r<r_E$ and $r_C<r<\infty$~\cite{GH}.
This may lead to the fact that the thermodynamic quantities of the
event horizon are not correct to the observer outside the
cosmological horizon. In this case we have to worry about the
Nariai transition at $r_+=r_0$ because of a forbidden region
($0\le r_E<r_0$) for the ESdS.

\section{Boundary CFT and Cardy-Verlinde formula}
The holographic principle means that the number of degrees of
freedom associated with the bulk gravitational dynamics is
determined by its boundary spacetime. The AdS/CFT correspondence
represents a realization of this principle~\cite{HOL}. For a
strongly coupled CFT with its AdS dual, one obtains the
Cardy-Verlinde formula~\cite{VER}. Indeed it is known that this
formula  holds for various kinds of asymptotically AdS spacetimes
including the TAdS black holes~\cite{CAI1}. Also it may hold for
asymptotically de Sitter spacetimes including the SdS black hole
and TdS spaces~\cite{CAI2}.  However, this formula is closely
related to the Hawking-Page transition~\cite{CMu,myungjmpa}. In
this section, we clarify the connection between Cardy-Verlinde
formula and Hawking-Page transition.  The boundary spacetimes for
the (E)CFT are defined through the A(dS)/CFT
correspondences~\cite{witten} \beqa \label{bcft}
&&ds^2_{CFT}=\lim_{r \to
\infty}\fr{R^2}{r^2}ds^2_{TAdS}=-\fr{R^2}{\ell^2}dt^2 +R^2
d\Sigma^2_k,\\
&&ds^2_{ECFT}=\lim_{r \to \infty}\fr{R^2}{r^2}
ds^2_{TdS}=\fr{R^2}{\ell^2} dt^2 +R^2 d\Sigma^2_k,\\
 &&
ds^2_{ECFT}=\lim_{r \to \infty}
\fr{R^2}{r^2}ds^2_{SdS}=\fr{R^2}{\ell^2}dt^2 +R^2 d\Omega^2_3.
\eeqa From the above,  the relation between the five-dimensional
bulk and four-dimensional boundary quantities is given by
$E_{CFT}=(\ell/R)E$ and $ T_{CFT}=(\ell/R)T_H$ where the size of
boundary space $R$ satisfies $T_{CFT}>1/R$. As is expected, we
obtain the same entropy : $S_{CFT}=S$. We note that the boundary
system at high temperature is described by the CFT-radiation
matter with the equation of state $p=E_{CFT}/3V_3$. Then  the
Casimir energy is given by $E_c=3(E_{CFT}+pV_3-T_{CFT}S_{CFT})$.
We find the boundary thermal quantities as functions of
$\hat{r}=r_{E/C}/\ell$ ~\cite{myungcqg} \beqa
&&E_{CFT}^{TAdS}=\fr{3V_3\kappa\hat{r}^2(\hat{r}^2+1)}{R},~~T_{CFT}^{TAdS}=\fr{1}{2\pi
 R}\Bigg[2\hat{r}+\frac{k}{\hat{r}}\Bigg],~~E_c^{TAdS}=k \fr{6
V_3 \kappa\hat{r}^2}{R},\\
&&E_{CFT}^{TdS}=\fr{3V_3\kappa\hat{r}^2(\hat{r}^2+1)}{R},~~T_{CFT}^{TdS}=\fr{1}{2\pi
 R}\Bigg[2\hat{r}-\frac{k}{\hat{r}}\Bigg],~~E_c^{TdS}=-k \fr{6
V_3 \kappa\hat{r}^2}{R},\\
 &&E_{CFT}^{E/C}=\pm \fr{3V_3\kappa\hat{r}^2(\hat{r}^2+1)}{R},~~
 T_{CFT}^{E/C}= \fr{1}{2\pi
 R}\Bigg[\mp2\hat{r}\pm \frac{1}{\hat{r}}\Bigg],~~E_c^{E/C}=\pm \fr{6
V_3 \kappa\hat{r}^2}{R} \eeqa with $\kappa=\ell^3/16 \pi G_5$.

Concerning the A(dS)/CFT correspondences, we remind the reader
that the boundary CFT energy ($E_{CFT}$) should be positive in
order for it to make sense. However, one
 finds  that $E_{CFT}^{CSdS}<0$ for the cosmological horizon of the SdS.
 It suggests that the dS/CFT correspondence is not realized  for this case.
 Also the Casimir energy ($E_c$) is related to the central charge
 of the corresponding CFT. If it is negative, one  may obtain
 a non-unitary CFT. In this sense HAdS, STdS, and CSdS cases may provide
 non-unitary conformal field theories. Furthermore, for FAdS and
 FTdS, there exist the forbidden region which gives rise to
 difficulties to define the temperature and the unstable black
 hole. Hence we guess that there is no realization of the (A)dS/CFT
 correspondences for Ricci-flat horizon.

  In order to find  a further implication of our study on dS/CFT
 correspondence, we discuss the Hawking-Page transition on the
 boundary system~\cite{myungjmpa}.  This corresponds to the transition between the confining and
 deconfining phase for ${\cal N}=4$ Super Yang Mills gauge theory as the CFT dual to SAdS (SAdS dual).
For this purpose we introduce the on-shell free energy defined by
$F_{CFT}(\hat{r})=E_{CFT}-T_{CFT}S_{CFT}$ \beq
F_{CFT}^{TAdS}=-\fr{V_3\kappa\hat{r}^2(\hat{r}^2-k)}{R},~F_{CFT}^{TAdS}=-\fr{V_3\kappa\hat{r}^2(\hat{r}^2+k)}{R},
~F_{CFT}^{TAdS}=\pm\fr{V_3\kappa\hat{r}^2(\hat{r}^2+1)}{R} \eeq
and off-shell free energy defined by \beq
F^{off}_{CFT}(\hat{r},T)=E_{CFT}-T S_{CFT} \eeq with the
temperature of the heat reservoir $T$. This corresponds to the
generalized free energy for the Landau-description of CFT phase
transition in Ref.\cite{CMu}.

 We guess that there is no  the confining/deconfining transition for HAdS
 and STdS duals because of $E_c<0$.  Also this transition  may not occur for CSdS dual because
of $E_c<0$ and $E_{CFT}<0$.  We show that the
confining/deconfining transition is possible to occur for HTdS
dual  by using the Landau-description of the phase
transition~\cite{CMu}. Finally, it seems that the dS/CFT
correspondence is not  realized for the SdS black hole because it
contains a black hole inside the cosmological horizon. Even for
 the Nariai transition from the black hole to de Sitter space at
$T=0$, it is not easy to see whether the confining/deconfining
transition occurs on its  CFT side.

Finally,  one finds the Cardy-Verlinde formulae which show the
exact relation between entropy $S_{CFT}$  and energy $E_{CFT}$ for
two cases \footnote{ For ESdS case, we have $S_{CFT}=(2 \pi R/3)
\sqrt{E_c(2E_{CFT}-E_c)}$. However, one could not convert it into
Eq.(\ref{cav1}) because there is no Hawking-Page transition at
$T=T_c$.}
only \beqa \label{cav1} && SAdS~ :~~S_{CFT}=\beta_c \sqrt{E_c(2E_{CFT}-E_c)} , \\
\label{cav2}&& HTdS~:~~S_{CFT}=\beta_c \sqrt{E_c(2E_{CFT}-E_c)}
\eeqa with the inverse of critical temperature $\beta_c=1/T_c= 2
\pi R/3$ determined by $F^{SAdS/HTdS}_{CFT}=0$.

\section{Discussion}
\begin{table}
 \caption{Summary of phase transitions
 for TAdS, TdS and SdS. Here $\cdot$(/) represent the continuous
 (discontinuous) sign changes.  HPT denotes Hawking-Page phase
 transition. A(d)-C means the A(dS)/CFT correspondences. }
 \begin{tabular}{|c|c|c|c|c|c|c|c|}
 \hline
  system & $r_u$ & $T_{H}$ &$E(r_+=r_1)$ & $C(r_+=r_0)$& $F(r_+=r_1)$ & HPT& A(d)-C  \\ \hline
  HAdS &N/A& N/A& $-\cdot+$& N/A &$-$& no &no \\ \hline
FAdS &0&+& +& +& $-$& no &no\\ \hline
 SAdS &+&+& +& $-\infty/\infty$& $+\cdot-$& yes &yes \\ \hline
 STdS &N/A&N/A&$-\cdot+$&  N/A &$-$ & no &no\\ \hline
FTdS &0&+ &+&  +&$-$ & no &no\\ \hline
 HTdS &+&+&+&$-\infty/\infty$& $+\cdot-$&yes &yes \\ \hline
ESdS-CSdS &+&+&$+/-(r_0)$& $-\cdot+(r_0)$ &$+/-(r_0)$ & no &no\\
\hline
 \end{tabular}
 \end{table}

We summarize our results in Table 1.
 For SAdS and
HTdS, the specific heat has a pole at $r_+=r_0$, while the free
energy is maximal at $r_+=r_0$ and  zero $r_+=r_1$. Here we have
two saddle points: a  small, unstable black hole and a large,
stable black hole. At $T=T_1$, we find the Hawking-Page phase
transition for SAdS and HTdS only. The transition point is at
$r_+=r_1$. Also the A(dS)/CFT correspondences are realized for
these cases.

On the other hand, for FADS and FTdS with  Ricci-flat horizon,
their specific heats are  monotonically increasing functions,
while  the free energies are monotonically decreasing functions.
We confirm that there is  no  Hawking-Page transition (A(dS)/CFT
correspondences~\cite{BIR}. However, for $T>0$, there may be  a
transition from a zero mass background to a black hole (de Sitter
space) by an off-shell process~\cite{myung3}. Even for taking the
AdS soliton background~\cite{SSW}, the situation is similar to the
case  here because there is no  small, unstable black hole.

In the cases of HAdS and STdS, there exist  forbidden regions.
Hence, we are unable to analyze the Hawking-Page transition  by
making use of the specific heat and free energy. We think that the
A(dS)/CFT correspondences are not realized for these.

Finally,  we mention the phase transition for the SdS. At $T=0$,
 the Nariai transition takes place from the black hole to de Sitter
space. At this temperature, the position of $r_+=r_0$ is not only
the  transition point but  the location of a stable cosmological
horizon. This is where the Nariai black hole is formed. However,
one cannot identify its dS/CFT correspondence.  On the other hand,
if one considers $0 \le r_E <r_0$ to be a forbidden region, there
is no Nariai transition at $T=0$.

\section*{Acknowledgments}
The author thanks Brian Murray  for reading the manuscript. This
work was in part  supported by the Korea Research Foundation Grant
(KRF-2005-013-C00018) and the SRC Program of the KOSEF through the
Center for Quantum Spacetime (CQUeST) of Sogang University with
grant number R11-2005-021-03001-0.

\end{document}